\documentstyle[lprocl,11pt,epsfig,psfig]{article}

\def\Journal#1#2#3#4{{#1} {\bf #2}, #3 (#4)}
\def\NCA{\em Nuovo Cimento}

\def\PRD{{\em Phys. Rev.} D}
\def\PR{{\em Phys. Rev.}}

\def\be{\begin{equation}}
\def\ee{\end{equation}}
\def\bea{\begin{eqnarray}}
\def\eea{\end{eqnarray}}
\newcommand{\nc}{\newcommand}
\nc{\noi}{\noindent}
\nc{\non}{\nonumber}
\nc{\s}[1]{\not \! #1}
\nc{\bb}{\bibitem}
\nc{\lf}{\left}
\nc{\r}{\right}
\nc{\mb}[1]{\makebox[#1]{}}
\nc{\pa}{\partial}
\nc{\sA}{\not \! \! A}
\nc{\newsec}[1]{\section{#1}\mb{0.5cm}}
\nc{\h}{\frac{1}{2}}
\nc{\ra}{\rightarrow}
\nc{\la}{\leftarrow}
\def\mathunderaccent#1{\let\theaccent#1\mathpalette\putaccentunder}
\def\putaccentunder#1#2{\oalign{$#1#2$\crcr\hidewidth
\vbox to.2ex{\hbox{$#1\theaccent{}$}\vss}\hidewidth}}

\nc{\ti}{\mathunderaccent\tilde}
\nc{\sq}{\sqrt{q^2}}

\nc{\bfig}{\begin{figure}}
\nc{\efig}{\end{figure}}
\nc{\half}{\frac{1}{2}}
\nc{\qtr}{\frac{1}{4}}
\nc{\halfP}{\frac{P}{2}}
\nc{\qtrP}{\frac{P^2}{4}}
\nc{\metric}[2]{\eta_{#1#2}}
\nc{\dpsq}[2]{\partial_\mu #1 \partial^\mu #2}
\nc{\ul}{\underline}
\nc{\intall}{\int_{-\infty}^{\infty}}
\nc{\inthalf}{\int_{0}^{\infty}}
\nc{\fullK}{\sum_{ch}\int_{\Omega}d\vec{\xi}\inthalf d\gamma
   \frac{\rho_{ch}(\gamma,\vec{\xi})}{\gamma-(a_{ch}q^2+b_{ch}p\cdot q + c_{ch}
   p^2+d_{ch}P^2 +e_{ch}q\cdot P + f_{ch}p\cdot P) - i\epsilon}}
\nc{\Kintch}{\frac{\rho_{ch}(\gamma,\vec{\xi})}{\gamma-(a_{ch}q^2+
   b_{ch}p\cdot q + c_{ch}p^2+d_{ch}P^2 +e_{ch}q\cdot P + f_{ch}p\cdot P) 
   - i\epsilon}}
\nc{\Kint}{\frac{\rho(\gamma,\vec{\xi})}{\gamma-(aq^2+ bp\cdot q + cp^2
   +dP^2 +eq\cdot P + fp\cdot P) - i\epsilon}}
\nc{\solid}[1]{{\cal Y}_{l}^{m}(#1)}
\nc{\solidq}{{\cal Y}_{l}^{m}(\Lambda^{-1}(P)q)}
\nc{\solidp}{{\cal Y}_{l}^{m}(\Lambda^{-1}(P)p)}
\nc{\loopint}{\int\frac{d^4q}{(2\pi)^4 i}}
\nc{\loopintprime}{\int\frac{d^4q'}{(2\pi)^4 i}}
\nc{\propone}[1]{\frac{1}{#1^2-m^2+i\epsilon}}
\nc{\proptwo}[1]{\frac{1}{(#1)^2-m^2+i\epsilon}}
\nc{\propsubone}[2]{\frac{1}{#2^2-m_{#1}^2+i\epsilon}}
\nc{\propsubtwo}[2]{\frac{1}{(#2)^2-m_{#1}^2+i\epsilon}}
\nc{\JH}{\frac{J}{H}}
\nc{\FH}{\frac{F}{H}}
\nc{\zb}{\bar{z}}
\nc{\ab}{\bar{\alpha}}
\nc{\gb}{\bar{\gamma}}
\nc{\at}{\tilde{\alpha}}
\nc{\zt}{\tilde{z}}
\nc{\gt}{\tilde{g}}
\nc{\vertex}[1]{\Gamma^{[l,m]}(#1,P)}
\nc{\ie}{i\epsilon}

\begin{document}

\title{COVARIANT SOLUTIONS OF THE BETHE-SALPETER EQUATION}

\author{ A.G. WILLIAMS, K. KUSAKA AND K.M. SIMPSON}

\address{Department of Physics and Mathematical Physics,\\
University of Adelaide, 
SA 5005, Australia}


\maketitle\abstracts{
There is a need for covariant solutions of bound state equations
in order to construct realistic QCD based models of mesons and baryons.
Furthermore, we ideally need to know the structure of these bound states
in all kinematical regimes, which makes a direct solution in Minkowski
space (without any 3-dimensional reductions) desirable.  The
Bethe-Salpeter equation (BSE) for bound states in scalar theories is
reformulated and solved for arbitrary scattering kernels in terms of a
generalized spectral representation directly in Minkowski space
(K. Kusaka et al., PRD 51, 7026 '95).  This differs from the conventional
Euclidean approach, where the BSE can only be solved in ladder approximation
after a Wick rotation. }

\section{Introduction}
The Bethe-Salpeter Equation (BSE)~\cite{BS} describes the 2-body component of 
bound-state structure relativistically and in the language of Quantum
Field Theory. 
It has applications in, for example, calculation of electromagnetic
form factors of 2-body bound states and in studies of relativistic 2-body bound
state spectra and wavefunctions.

BSEs have been solved 
previously for separable kernels and for ladder scattering kernels.
Solutions for BSEs have also been obtained for QCD-based models of meson
structure in Euclidean space; 
these solutions must be 
analytically continued to Minkowski space. If we solve the BSE with a 
non-ladder scattering kernel or dressed propagators for the constituent
bodies, the validity of this procedure (known as the Wick rotation~\cite
{Wick}) is highly non-trivial, and so direct solution in Minkowski space 
without unnecessary approximations is preferable. Here we outline such a 
method for scalar theories, based on
the Perturbation Theoretic Integral Representation (PTIR) of
Nakanishi~\cite{PTIR}. 

The PTIR is a generalisation of the spectral 
representation for 2-point Green's functions to $n$-point functions;
for a particular
renormalised $n$-point function, the PTIR is an integral representation of
the corresponding infinite sum of Feynman graphs with $n$
external legs.

Our approach involves using the PTIR to transform the equation for the proper 
bound-state vertex, which is an integral equation involving complex 
distributions, into a real integral equation. 
This equation may then be solved numerically
for an arbitrary scattering kernel~\cite{kw}. We have to date applied this
formalism to the case of a ladder kernel, as well as the so-called
``dressed ladder'' kernel in which we add self-energy corrections to
the propagator of the exchanged particle. These results have been verified
by direct comparison 
to those obtained in Euclidean space by previous authors~\cite{lm}.

An example of one such scalar theory to
which our formalism may be applied is the $\phi^2\sigma$ model,
which has a Lagrangian density
\be
{\cal L} = \half(\dpsq{\phi}{\phi} - m^{2} \phi^{2}) + \half(\dpsq
{\sigma}{\sigma} - m_{\sigma}^{2}\sigma^2) - g\phi^2\sigma,
\label{lagrangian}
\ee
\noi where $g$ is the $\phi$-$\sigma$ coupling constant.

\section{Formalism and PTIR}

The Bethe-Salpeter equation in momentum space for a scalar-scalar bound 
state with scalar exchange is 
\be
\Phi(p,P)=-D(p_{1})D(p_{2})\int \frac{d^4q}{(2\pi)^4}\Phi(q,P)K(p,q;P),
\label{BSalg}
\ee
\noi where $\Phi$ is the Bethe-Salpeter (BS) amplitude, and where $K$
is the scattering kernel, which contains information about the interactions
between the constituents of the bound state. We may also write this in terms
of the bound state vertex $\Gamma$ as
\be
\Gamma(p,P)=\int \frac{d^4q}{(2\pi)^4}D(p_1)D(p_2)K(p,q;P)\Gamma(q,P),
\label{BSalg2}
\ee

In Eq.~(\ref{BSalg}), $p_{i}$ is the four-momentum of the $i^{\rm th}$ 
constituent. We also define $p\equiv\eta_{2}p_{1}-\eta_{1}p_{2}$, which is 
the relative four-momentum of the two constituents, and $P\equiv p_{1}+p_{2}$ 
is the total 
four-momentum of the bound-state. The real positive numbers $\eta_{i}$ are 
arbitrary, with the only constraint being that $\eta_{1}+\eta_{2}=1$. 
We will use $\eta_{1}=1/2=\eta_{2}$, which is a convenient choice when the
constituents have equal mass.

In order to convert the BSE into a real integral equation, we will need to
use the PTIR for both the
proper bound-state vertex and the scattering kernel~\cite{kw}. The bound-state
vertex may be represented as
\be
\Gamma(q,P)=\inthalf d\alpha\int_{-1}^{1}dz\:
\frac{\rho_n(\alpha,z)}{[\alpha+m^2-(q^2+zq\cdot P+\qtrP)-i\epsilon]^n}.
\label{basicvertex}
\ee
The weight function $\rho_n(\alpha,z)$ of the vertex has support
only for a finite region of the space spanned by the parameters $\alpha$
and $z$. 

The representation for the vertex in Eq.~(\ref{basicvertex}) is for $s$-wave
bound states. It is straightforward to generalise our 
arguments to higher partial waves~\cite{kw,kkt}.

We have introduced a dummy parameter $n$, which will be of use in our
numerical work since larger values of $n$ produce smoother weight 
functions. The fact that $n$ is arbitrary can be seen by integrating
by parts with respect to $\alpha$; in this way weight functions for
different values of $n$ may be connected~\cite{kw}. 

We may use the PTIR for the bound-state vertex to derive the PTIR for the
Bethe-Salpeter amplitude, since the two are related via
\be
  \Phi(p,P)=iD(\halfP+p)i\Gamma(p,P)iD(\halfP-p).
\label{phigamma}
\ee
We proceed by absorbing the two free scalar propagators into the expression
for the vertex, Eq.~(\ref{basicvertex}), using Feynman 
parametrisation. After some algebra we obtain for the BS amplitude
\be
\Phi(p,P)=-i\int_{-1}^{1}dz
\intall d\alpha \frac{\varphi_n(\alpha,z)}{\left[m^2+\alpha-
(p^2+zp\cdot P +\qtrP)-i\epsilon\right]^{n+2}},
\label{pwBSE}
\ee

To include the most general form of the scattering kernel in our derivation,
we use the PTIR for the kernel:
\bea
-iK(p,q;P)&=&\sum_{\rm ch}\inthalf d\gamma\int_{\Omega}d\vec{\xi}\:\non\\
& &\frac{\rho_{\rm ch}(\gamma,\vec{\xi})}{\gamma-(a_{\rm ch}q^2+b_{\rm ch}p
\cdot q+c_{\rm ch}p^2+d_{\rm ch}P^2+e_{\rm ch}q\cdot P+f_{\rm ch}p\cdot P)
-i\epsilon}, \non\\
& &
\label{generalkernel}
\eea
\noi where the kernel parameters $\{a_{\rm ch},\ldots,f_{\rm ch}\}$ are 
linear combinations of the Feynman parameters $\{\xi_1,\ldots,\xi_6\}$,
and $\Omega$ denotes the region of integration which is that imposed by
the constraint $\sum_{i=1}^6 \xi_i=1$. 
Here we have defined, similarly to before, $q\equiv (q_1-q_2)/2$. The 
expression for the kernel contains a sum over three different channels,
labelled by $st$, $tu$ and $us$.

\section{Derivation of Equations for Scalar Models}
We use the PTIR form of the kernel and vertex in Eq.~(\ref{BSalg2}),
along with Feynman parametrisation, to derive a real, two-dimensional 
integral equation for the vertex weight function $\rho(\alpha,z)$. 
We use bare constituent propagators, and using the PTIR uniqueness
theorem~\cite{PTIR}, obtain the form of the equation that can be solved
numerically,
\be
\frac{1}{\lambda}\frac{\rho_n(\ab,\zb)}{\ab^n}=\inthalf d\alpha\int_{-1}^{1}
dz\: {\cal K}_n(\ab,\zb;\alpha,z)\:\frac{\rho_n(\alpha,z)}{\alpha^n}
\label{inteqn}
\ee

We have defined here an eigenvalue $\lambda\equiv g^2/(4\pi)^2$, where
$g$ is the $\phi$--$\sigma$ coupling.
We solve Eq.~(\ref{inteqn}) numerically by iteration to convergence for the 
modified weight function $\rho_n(\ab,\zb)/\ab^n$. For the sake of brevity,
we will not write down the explicit form of the kernel function ${\cal K}$
here. Note that ${\cal K}$ should not be confused with the scattering
kernel, $K$, discussed earlier.

\begin{figure}[hbt]
  \centering{\
     \epsfig{angle=0,figure=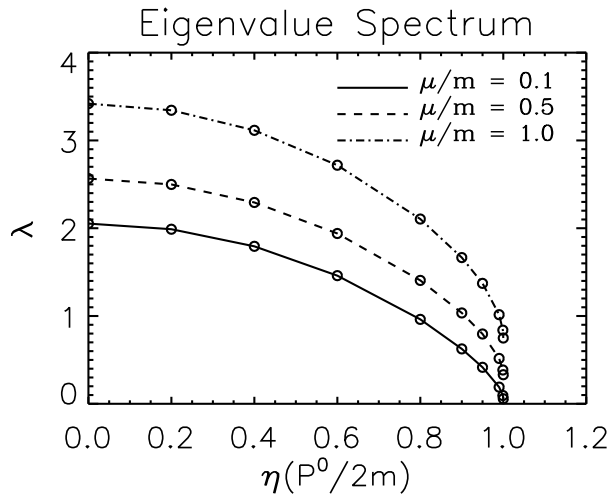,height=7.0cm} }
\protect\parbox{110mm}{\caption{Eigenvalue spectrum for the scalar BSE. 
Here 
\mbox{$\mu/m$}
is the ratio of the exchange mass to the constituent mass, and 
\mbox{$\eta$}
is 
a parameter which describes the ``fraction of binding'': 
\mbox{$\eta\equiv({P^2/4m^2})^{\frac{1}{2}}$},
which in the rest frame of the bound state 
reduces to 
\mbox{$P^0/2m$}.
}
\label{spec}}
\end{figure}

\begin{figure}[h]
  \centering{\
     \epsfig{angle=0,figure=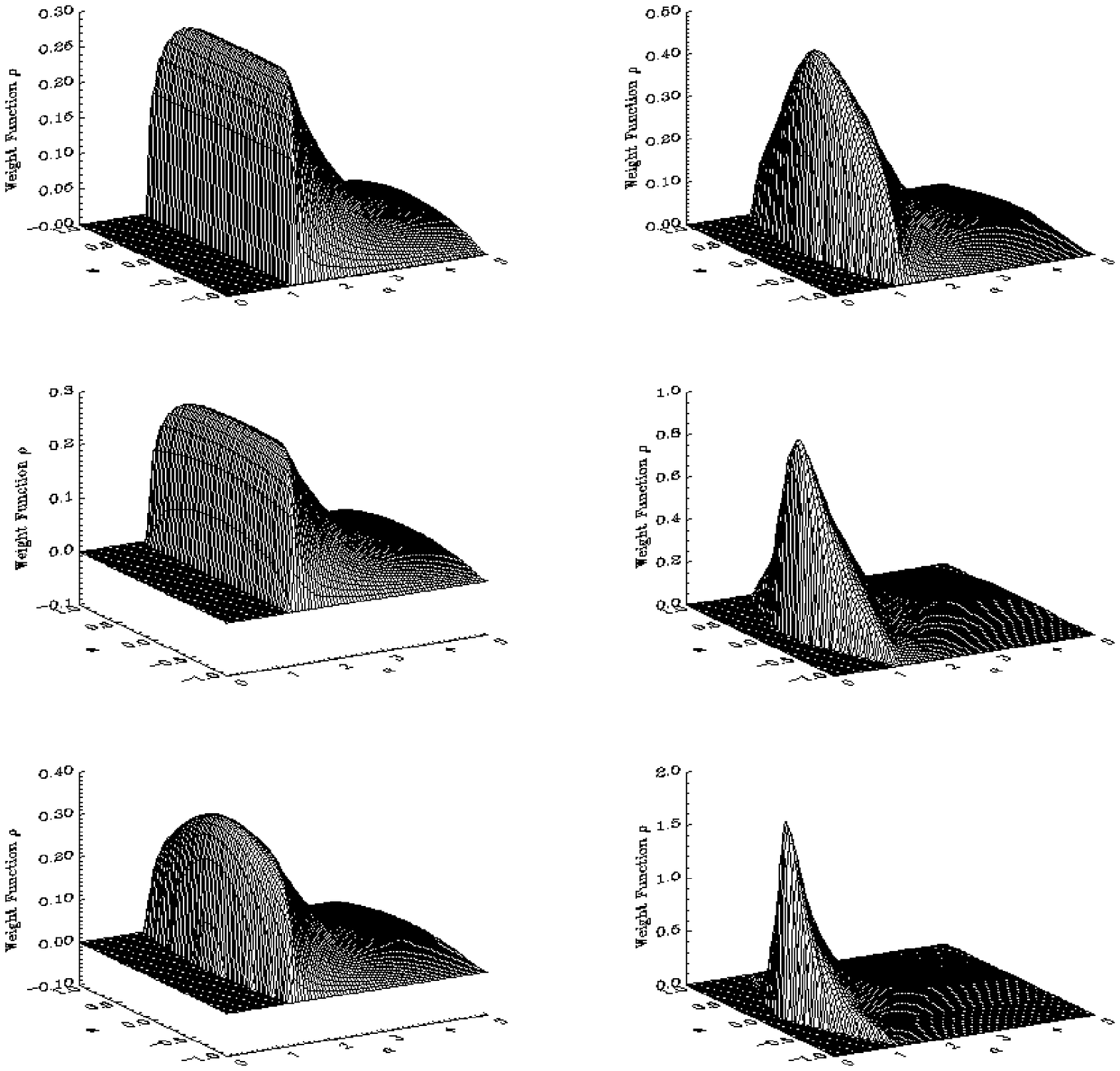,height=14.5cm} }
\protect\parbox{130mm}{\caption{Development of the weight function as the 
bound state mass increases towards threshold. The left column contains the
cases \mbox{$\eta$}=0,0.2 and 0.6, while the right hand column contains the
cases \mbox{$\eta$}=0.8,0.95 and 0.99.}
\label{weight}}
\end{figure}

\begin{figure}[h]
  \centering{\
     \epsfig{angle=0,figure=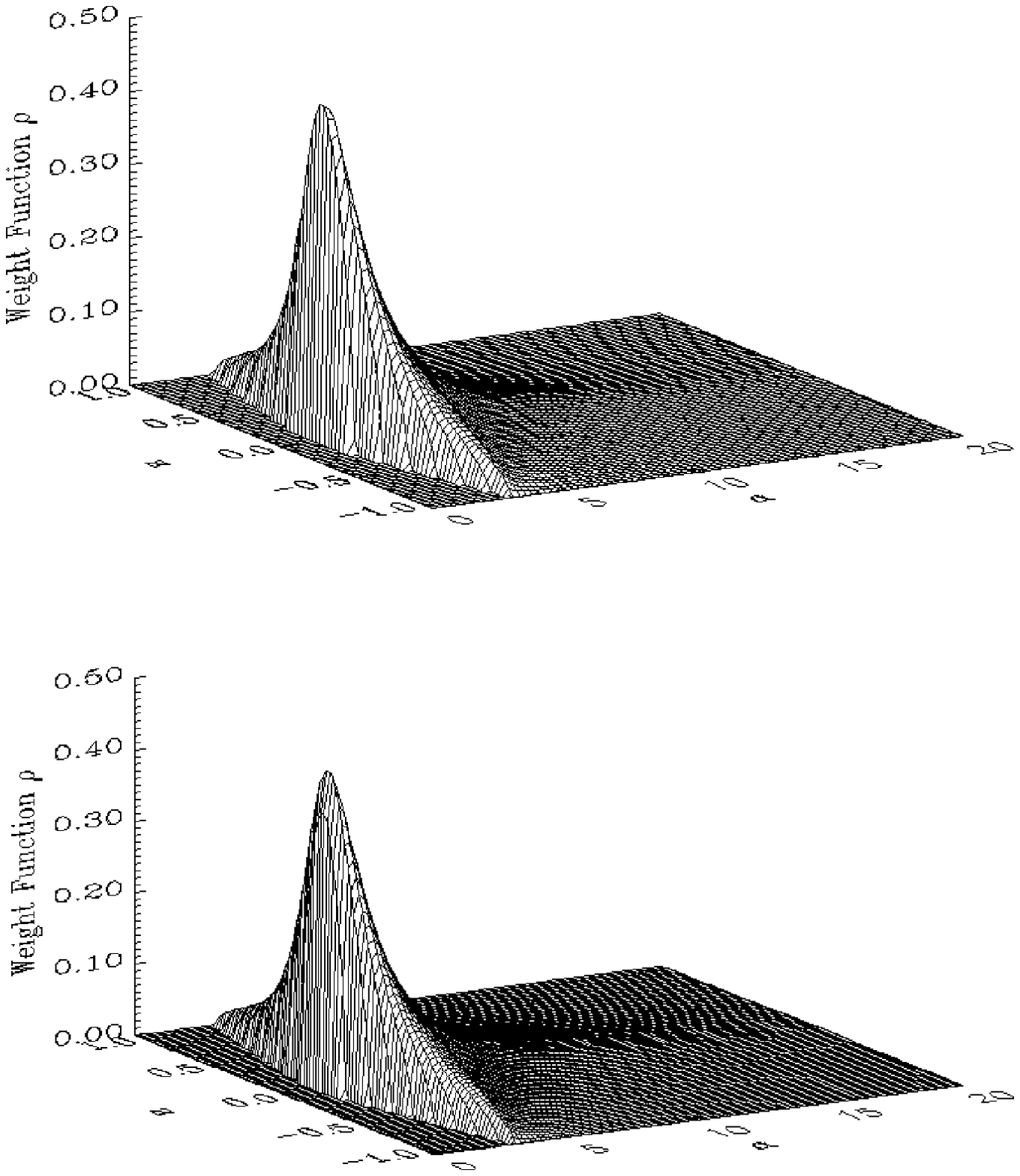,height=12.5cm} }
\protect\parbox{130mm}{\caption{Comparison of the weight functions for
a ladder kernel (top figure) and a dressed ladder kernel (bottom figure).}
\label{compare}}
\end{figure}

As a test of our implementation of the formalism presented here, we have solved
Eq.~(\ref{inteqn}) with $n=2$ for a ladder kernel and for a dressed ladder
kernel where we add a one-loop self-energy correction to the propagator
of the exchange ($\sigma$) particle. Our eigenvalues agree with those obtained
by Linden and Mitter~\cite{lm} in Euclidean space to typically 1 part in 10$^4$
or better, for moderate numbers of grid points. We plot the eigenvalue
spectrum for ladder exchange for various values of the exchange mass $\mu$
in Fig.~\ref{spec}. Fig.~\ref{weight} illustrates the development of the 
modified weight function $\rho_2(\alpha,z)/\alpha^2$ obtained by solving the
ladder BSE (with $\mu/m=0.5$) as we increase the value of the 
bound state mass squared, $P^2$, towards the instability threshold where 
$P^2=4m^2$. As $P^2$ increases towards 
threshold, the binding becomes weaker, {\em i.e.}, $g$ decreases. Finally, 
in Fig.~\ref{compare} we compare the weight functions for the ladder and 
dressed ladder cases (here the pole in the $\sigma$--propagator is located 
at $\mu/m=1.0$) when the binding is quite weak 
($\sqrt{P^2/4m^2}=0.99$). We found that the one-loop correction enhances 
the binding very slightly (by a few percent or smaller), and does not have a 
significant effect on the weight function~\cite{kkt}, as is borne out by 
Fig.~\ref{compare}.

\section{Conclusions and Outlook}
We have formulated the BSE in Minkowski space for a completely general
scattering kernel. This has been solved numerically for the ladder and
dressed ladder kernels; both sets of results agree to high accuracy with 
results using more traditional Euclidean space methods.

We are currently implementing a kernel code which will enable us to
numerically solve the scalar BSE for a completely general scalar kernel. 
As a first application of this code, we will solve for a ``generalised 
ladder'' kernel~\cite{kkt}. The Wick rotation is invalid in this case
and so solution of the BSE in Minkowski space is essential in this instance.

To study problems of more general interest, it will be essential to extend the
PTIR to theories containing fermions, and to find a way of incorporating
confinement and derivative coupling into our approach. This will enable
us to carry out studies of mesons in QCD, for example.

\end{document}